\magnification=\magstephalf
\font\t=cmcsc10 at 13 pt
\font\tt=cmcsc10
\font\n=cmcsc10
\font\foot=cmr9
\font\abs=cmr8
\font\babs=cmbx8
\centerline{\t On the Gravitational Field of a Mass Point}
\smallskip
\centerline{\t according to Einstein's Theory \footnote{\dag}
{\foot{Sitzungsberichte der K\"oniglich Preussischen Akademie
der Wissenschaften zu Berlin, Phys.-Math. Klasse 1916, 189-196.}}}\bigskip
\centerline{\n by K. Schwarzschild}\bigskip
\centerline{(Communicated January 13th, 1916 [see above p.
42].)}\par\vbox to 0.6 cm {}
\centerline{\tt translation\footnote{\ddag}{\foot The valuable
advice of D.-E. Liebscher is gratefully acknowledged.}
and foreword by}\smallskip
\centerline{S. Antoci\footnote{$^*$}{\foot Dipartimento di Fisica
``A. Volta'', Universit\`a di Pavia, Via Bassi 6 - 27100 Pavia
(Italy).} and A. Loinger\footnote{$^{**}$}{\foot Dipartimento di Fisica,
Universit\`a di Milano, Via Celoria 16 - 20133 Milano
(Italy).}}\medskip
{\babs Foreword.} {\abs This fundamental memoir contains
the ORIGINAL form of the solution of Schwarzschild's problem.
It is regular in the whole space-time, with the only exception
of the origin of the spatial co-ordinates; consequently, it
leaves no room for the science fiction of the black holes.
(In the centuries of the decline of the Roman Empire people said:
``Graecum est, non legitur''...).}\par
\vbox to 0.6 cm {}

\S 1. In his work on the motion of the perihelion of Mercury (see
Sitzungsberichte of November 18th, 1915) Mr. Einstein has posed
the following problem:\par
Let a point move according to the prescription:

$$\left\{\eqalign{\delta&\int{ds}=0,\cr
&where\cr
ds&=\sqrt{\Sigma g_{\mu\nu}dx_\mu
dx_\nu}~~\mu,\nu=1,2,3,4,\cr}\right.\eqno(1)$$
where the $g_{\mu\nu}$ stand for functions of the variables $x$, and
in the variation the variables $x$ must be kept fixed at the
beginning and at the end of the path of integration. In short, the
point shall move along a geodesic line in the manifold
characterised by the line element $ds$.\par
The execution of the variation yields the equations of motion of the
point:

$${d^2x_\alpha\over ds^2}={\sum_{\mu,\nu}}~\Gamma^\alpha_{\mu\nu}
{dx_\mu\over ds}{dx_\nu\over ds},~~\alpha,\beta=1,2,3,4,\eqno(2)$$
where

$$\Gamma^\alpha_{\mu\nu}=-{1\over 2}\sum_\beta g^{\alpha\beta}
\bigg({\partial g_{\mu\beta}\over \partial x_\nu}+
{\partial g_{\nu\beta}\over \partial x_\mu}
-{\partial g_{\mu\nu}\over \partial x_\beta}\bigg),\eqno(3)$$
and the $g^{\alpha\beta}$ stand for the normalised minors associated to
$g_{\alpha\beta}$ in the determinant $\vert g_{\mu\nu}\vert$.\par
According to Einstein's theory, this is the motion of a
massless point in the gravitational field of a mass at
the point $x_1=x_2=x_3=0$, if the ``components of the
gravitational field'' $\Gamma$ fulfil everywhere, with the
exception of the point $x_1=x_2=x_3=0$, the ``field equations''

$$\sum_\alpha{\partial \Gamma^\alpha_{\mu\nu}\over {\partial
x_\alpha}}
+\sum_{\alpha\beta}~\Gamma^\alpha_{\mu\beta
}\Gamma^\beta_{\nu\alpha}=0,\eqno(4)$$
and if also the ``equation of the determinant''

$$\vert g_{\mu\nu}\vert=-1\eqno(5)$$
is satisfied.\par
The field equations together with the equation of the
determinant have the fundamental property that they preserve their
form under the substitution of other arbitrary variables in lieu
of $x_1$, $x_2$, $x_3$, $x_4$, as long as the determinant of the
substitution is equal to $1$.\par
Let $x_1$, $x_2$, $x_3$ stand for rectangular co-ordinates, $x_4$ for the
time; furthermore, the mass at the origin shall not change with time, and the
motion at infinity shall be rectilinear and uniform. Then,
according to Mr. Einstein's list, {\it loc. cit.} p. 833, the following
conditions must be fulfilled too:\smallskip
\item{1.} All the components are independent of the time $x_4$.
\item{2.} The equations $g_{\rho4}=g_{4\rho}=0$ hold exactly for
$\rho=1, 2, 3.$
\item{3.} The solution is spatially symmetric with respect to the
origin of the co-ordinate system in the sense that one finds again
the same solution when $x_1$, $x_2$, $x_3$ are subjected to an
orthogonal transformation (rotation).
\item{4.} The $g_{\mu\nu}$ vanish at infinity, with the exception
of the following four limits different from zero:

$$g_{44}=1,~~g_{11}=g_{22}=g_{33}=-1.$$
\par\noindent
{\it The problem is to find out a line element with
coefficients such that the field equations, the equation of the
determinant and these four requirements are satisfied.}\par
\S 2. Mr. Einstein showed that this problem, in first
approximation, leads to Newton's law and that the second
approximation correctly reproduces the known anomaly in the motion
of the perihelion of Mercury. The following calculation yields
the exact solution of the problem. It is always pleasant to avail
of exact solutions of simple form. More importantly, the
calculation proves also the uniqueness of the solution, about
which Mr. Einstein's treatment still left doubt, and which could have
been proved only with great difficulty, in the way shown below,
through such an approximation method. The following lines
therefore let Mr. Einstein's result shine with increased
clearness.\par
\S 3. If one calls $t$ the time, $x$, $y$, $z$, the rectangular
co-ordinates, the most general line element that satisfies the
conditions 1-3 is clearly the following:

$$ds^2=Fdt^2-G(dx^2+dy^2+dz^2)-H(xdx+ydy+zdz)^2$$
where $F$, $G$, $H$ are functions of $r=\sqrt{x^2+y^2+z^2}$.\par
The condition (4) requires: for $r=\infty: F=G=1, H=0$.\par
When one goes over to polar co-ordinates according to
$x=r\sin\vartheta\cos\phi,~y=r\sin\vartheta\sin\phi,~
z=r\cos\vartheta,$
the same line element reads:

$$\eqalign{ds^2&=Fdt^2
-G(dr^2+r^2d\vartheta^2+r^2sin^2\vartheta d\phi^2)-Hr^2dr^2\cr
&=Fdt^2-(G+Hr^2)dr^2-Gr^2(d\vartheta^2+sin^2\vartheta
d\phi^2).}\eqno(6)$$
Now the volume element in polar co-ordinates is equal to
$r^2\sin\vartheta drd\vartheta d\phi$, the functional determinant
$r^2\sin\vartheta$ of the old with respect to the new coordinates
is different from $1$; then the field equations would not remain in
unaltered form if one would calculate with these polar
co-ordinates, and one would have to perform a cumbersome transformation.
However there is an easy trick to circumvent this difficulty. One
puts:

$$x_1={r^3\over 3},~~x_2=-\cos\vartheta,~~x_3=\phi.\eqno(7)$$
Then we have for the volume element:
$r^2dr\sin\vartheta d\vartheta d\phi=dx_1dx_2dx_3.$ The new
variables are then {\it polar co-ordinates with the determinant 1}.
They have the evident advantages of polar co-ordinates for the
treatment of the problem, and at the same time, when
one includes also $t=x_4$, the field equations and the determinant
equation remain in unaltered form.\par
In the new polar co-ordinates the line element reads:

$$ds^2=Fdx_4^2-\bigg({G\over r^4}+{H\over r^2}\bigg)dx_1^2
-Gr^2\bigg[{dx_2^2\over{1-x_2^2}}+dx_3^2(1-x_2^2)\bigg],\eqno(8)$$
for which we write:

$$ds^2=f_4dx_4^2-f_1dx_1^2-f_2{dx_2^2\over {1-x_2^2}}
-f_3dx_3^2(1-x_2^2).\eqno(9)$$
Then $f_1$, $f_2=f_3$, $f_4$ are three functions of $x_1$ which
have to fulfil the following conditions:\par\smallskip

\item{1.} For $x_1=\infty:~f_1={1/
r^4}=(3x_1)^{-4/3},~f_2=f_3=r^2=(3x_1)^{2/3},~f_4=1$.
\item{2.} The equation of the determinant:
$f_1\cdot f_2\cdot f_3\cdot f_4=1$.
\item{3.} The field equations.
\item{4.} Continuity of the $f$, except for $x_1=0$.\par\smallskip
\S 4. In order to formulate the field equations one must first
form the components of the gravitational field corresponding to the
line element (9). This happens in the simplest way when one builds
the differential equations of the geodesic line by direct
execution of the variation, and reads out the components from these.
The differential equations of the geodesic line for the line
element (9) result from the variation immediately in the form:

$$\eqalign{0&=f_1{d^2x_1\over ds^2}+{1\over 2}{\partial f_4\over
\partial x_1}\bigg({dx_4\over ds}\bigg)^2
+{1\over 2}{\partial f_1\over
\partial x_1}\bigg({dx_1\over ds}\bigg)^2
-{1\over 2}{\partial f_2\over
\partial x_1}
\bigg[{1\over{1-x_2^2}}\bigg({dx_2\over ds}\bigg)^2
+(1-x_2^2){\bigg({dx_3\over{ ds}}\bigg)^2}\bigg]\cr
&\cr
0&={f_2\over{1-x_2^2}}{d^2x_2\over ds^2}+{\partial f_2\over
\partial x_1}{1\over{1-x_2^2}}{dx_1\over ds}{dx_2\over ds}
+{f_2x_2\over{(1-x_2^2)^2}}\bigg({dx_2\over ds}\bigg)^2
+f_2x_2\bigg({dx_3\over ds}\bigg)^2\cr
&\cr
0&=f_2(1-x_2^2){d^2x_3\over ds^2}+{\partial f_2\over
\partial x_1}(1-x_2^2){dx_1\over ds}{dx_3\over ds}
-2f_2x_2{dx_2\over ds}{dx_3\over ds}\cr
&\cr
0&=f_4{d^2x_4\over ds^2}+{\partial f_4\over
\partial x_1}{dx_1\over ds}{dx_4\over ds}.}$$
The comparison with (2) gives the components of the gravitational
field:

$$\eqalign{\Gamma^1_{11}&=-{1\over 2}{1\over f_1}{\partial f_1\over
\partial x_1},~~~\Gamma^1_{22}=+{1\over 2}{1\over f_1}{\partial f_2\over
\partial x_1}{1\over{1-x_2^2}},\cr
\Gamma^1_{33}&=+{1\over 2}{1\over f_1}{\partial f_2\over
\partial x_1}(1-x_2^2),\cr
\Gamma^1_{44}&=-{1\over 2}{1\over f_1}{\partial f_4\over
\partial x_1},\cr
\Gamma^2_{21}&=-{1\over 2}{1\over f_2}{\partial f_2\over
\partial x_1},~~~\Gamma^2_{22}=-{x_2\over{1-x_2^2}},~~~
\Gamma^2_{33}=-x_2(1-x_2^2),\cr
\Gamma^3_{31}&=-{1\over 2}{1\over f_2}{\partial f_2\over
\partial x_1},~~~\Gamma^3_{32}=+{x_2\over{1-x_2^2}},\cr
\Gamma^4_{41}&=-{1\over 2}{1\over f_4}{\partial f_4\over
\partial x_1}}$$
\centerline{(the other ones are zero).} Due to the rotational symmetry
around the origin  it is sufficient to write the field equations
only for the equator ($x_2=0$); therefore, since they will be
differentiated only once, in the previous expressions it is
possible to set everywhere since the beginning $1-x_2^2$ equal
$1$. The calculation of the field equations then gives

$$\eqalign{&a)~{\partial\over{\partial x_1}}\bigg({1\over f_1}
{\partial f_1\over{\partial x_1}}\bigg) ={1\over 2}\bigg({1\over
f_1} {\partial f_1\over{\partial x_1}}\bigg)^2 +\bigg({1\over f_2}
{\partial f_2\over{\partial x_1}}\bigg)^2 +{1\over 2}\bigg({1\over
f_4} {\partial f_4\over{\partial x_1}}\bigg)^2,\cr
&b)~{\partial\over{\partial x_1}}\bigg({1\over f_1} {\partial
f_2\over{\partial x_1}}\bigg) =2+{1\over{f_1f_2}}\bigg( {\partial
f_2\over{\partial x_1}}\bigg)^2,\cr &c)~{\partial\over{\partial
x_1}}\bigg({1\over f_1} {\partial f_4\over{\partial x_1}}\bigg)
={1\over{f_1f_4}}\bigg( {\partial f_4\over{\partial
x_1}}\bigg)^2.}$$
Besides these three equations the functions
$f_1$, $f_2$, $f_3$ must fulfil also the equation of the
determinant
$$d)~f_1f_2^2f_4=1,~i.~e.~~{1\over f_1} {\partial
f_1\over{\partial x_1}} +{2\over f_2} {\partial f_2\over{\partial
x_1}} +{1\over f_4} {\partial f_4\over{\partial x_1}}=0.$$
For now I neglect (b) and determine the three functions $f_1$,
$f_2$, $f_4$ from (a), (c), and (d). (c) can be transposed into the
form

$$c')~{\partial\over{\partial x_1}}\bigg({1\over f_4}
{\partial f_4\over{\partial x_1}}\bigg) ={1\over{f_1f_4}}
{\partial f_1\over{\partial x_1}} {\partial f_4\over{\partial
x_1}}.$$
This can be directly integrated and gives

$$c'')~={1\over f_4} {\partial f_4\over{\partial x_1}} =\alpha
f_1,~~~(\alpha~integration~constant)$$ The addition of (a) and
(c') gives $${\partial\over{\partial x_1}}\bigg({1\over f_1}
{\partial f_1\over{\partial x_1}} +{1\over f_4} {\partial
f_4\over{\partial x_1}}\bigg) =\bigg({1\over f_2} {\partial
f_2\over{\partial x_1}}\bigg)^2 +{1\over 2}\bigg({1\over f_1}
{\partial f_1\over{\partial x_1}} +{1\over f_4} {\partial
f_4\over{\partial x_1}}\bigg)^2.$$
By taking (d) into account it follows

$$-2{\partial\over{\partial x_1}}\bigg({1\over f_2} {\partial
f_2\over{\partial x_1}}\bigg) =3\bigg({1\over f_2} {\partial
f_2\over{\partial x_1}}\bigg)^2.$$
By integrating

$${1\over{{1\over f_2} {\partial f_2\over{\partial x_1}}}}={3\over
2}x_1+{\rho\over 2}~~~ (\rho~integration~constant)$$ or $${1\over
f_2}{\partial f_2\over{\partial x_1}} ={2\over{3x_1+\rho}}.$$
By integrating once more,

$$f_2=\lambda(3x_1+\rho)^{2/3}.~~~(\lambda~integration~constant)$$
The condition at infinity requires:
$\lambda=1$. Then
$$f_2=(3x_1+\rho)^{2/3}.\eqno(10)$$
Hence it results further from (c'') and (d)
$${\partial f_4\over\partial x_1}=\alpha f_1f_4={\alpha\over
f_2^2}={\alpha\over{(3x_1+\rho)^{4/3}}}.$$
By integrating while taking into account the condition at infinity

$$f_4=1-\alpha(3x_1+\rho)^{-1/3}.\eqno(11)$$
Hence from (d)

$$f_1={(3x_1+\rho)^{-4/3}
\over{1-\alpha(3x_1+\rho)^{-1/3}}}.\eqno(12)$$ As can be easily
verified, the equation (b) is automatically fulfilled by the
expressions that we found for $f_1$ and $f_2$.\par Therefore all the
conditions are satisfied apart from the {\it condition of continuity}.
$f_1$ will be discontinuous when

$$1=\alpha(3x_1+\rho)^{-1/3},~~~3x_1=\alpha^3-\rho.$$
In order that this discontinuity coincides with the origin,
it must be

$$\rho=\alpha^3.\eqno(13)$$
Therefore the condition of continuity relates in this way the
two integration constants $\rho$ and $\alpha$.\par The complete
solution of our problem reads now:

$$f_1={1\over R^4}{1\over{1-\alpha/R}}
,~~f_2=f_3=R^2,~~f_4=1-\alpha/R,$$
where the auxiliary quantity

$$R=(3x_1+\rho)^{1/3}=(r^3+\alpha^3)^{1/3}$$
has been introduced.\par {\it When one introduces these values of the
functions $f$ in the expression (9) of the line element and goes
back to the usual polar co-ordinates one gets the line element
that forms the exact solution of Einstein's problem}:

$$ds^2=(1-\alpha/R)dt^2-{dR^2\over{1-\alpha/R}}
-R^2(d\vartheta^2+\sin^2\vartheta
d\phi^2),~~R=(r^3+\alpha^3)^{1/3}.\eqno(14)$$
The latter contains only the constant $\alpha$ that depends on
the value of the mass at the origin.\par
\S 5. {\it The uniqueness of the solution} resulted
spontaneously through the present calculation. From what follows
we can see that it would have been difficult to ascertain the
uniqueness from an approximation procedure in the manner of
Mr. Einstein. Without the continuity condition it
would have resulted:

$$f_1={(3x_1+\rho)^{-4/3}\over{1-\alpha(3x_1+\rho)^{-1/3}}}
={(r^3+\rho)^{-4/3}\over{1-\alpha(r^3+\rho)^{-1/3}}}.$$
When $\alpha$ and $\rho$ are small, the series expansion up to
quantities of second order gives:
$$f_1={1\over r^4}
\bigg[1+{\alpha\over r}-4/3{\rho\over r^3}\bigg].$$
This expression, together with the corresponding expansions of $f_2$,
$f_3$, $f_4$, satisfies up to the same accuracy all the
conditions of the problem. Within this approximation the condition
of continuity does not introduce anything new, since discontinuities
occur spontaneously only in the origin. Then the two constants
$\alpha$ and $\rho$ appear to remain arbitrary, hence the problem
would be physically undetermined. The exact solution teaches that
in reality, by extending the approximations, the discontinuity does
not occur at the origin, but at
$r=(\alpha^3-\rho)^{1/3}$, and that one must set just
$\rho=\alpha^3$ for the discontinuity to go in the
origin. With the approximation in powers of $\alpha$ and $\rho$
one should survey very closely the law of the coefficients in
order to recognise the necessity of this link between $\alpha$ and
$\rho$.\par
\S 6. Finally, one has still to derive the {\it
motion of a point in the gravitational field}, the geodesic line
corresponding to the line element (14). From the three facts, that
the line element is homogeneous in the differentials and that its
coefficients do not depend on $t$ and on $\phi$, with the
variation we get immediately three intermediate integrals. If
one also restricts himself to the motion in the equatorial plane
($\vartheta=90^o,~d\vartheta=0$) these intermediate integrals
read:
$$(1-\alpha/R)\bigg({dt\over ds}\bigg)^2 -{1\over{1-\alpha/R}}
\bigg({dR\over ds}\bigg)^2-R^2\bigg({d\phi\over
ds}\bigg)^2=const.=h,\eqno(15)$$

$$R^2{d\phi\over ds}=const.=c,\eqno(16)$$

$$(1-\alpha/R){dt\over ds}=const.=1~~~(determination~of~the~time~
unit).\eqno(17)$$

\noindent From here it follows

$$\bigg({dR\over d\phi}\bigg)^2+R^2(1-\alpha/R)
={R^4\over c^2}[1-h(1-\alpha/R)]$$
or with $1/R=x$
$$\bigg({dx\over d\phi}\bigg)^2={{1-h}\over
c^2}+{{h\alpha}\over c^2}x-x^2+\alpha x^3.\eqno(18)$$
If one introduces the notations: ${c^2/h}=B$, ${(1-h)/h}=2A$,
this is identical to Mr. Einstein's equation (11),
{\it loc. cit.} and gives the observed anomaly of the perihelion of
Mercury.\par
Actually Mr. Einstein's approximation for the orbit goes into
the exact solution when one substitutes for $r$ the quantity

$$R=(r^3+\alpha^3)^{1/3}
=r\bigg(1+{\alpha^3\over r^3}\bigg)^{1/3}.$$
Since $\alpha/r$ is nearly equal to twice the square of the
velocity of the planet (with the velocity of light as unit),
for Mercury the parenthesis differs from $1$ only for quantities
of the order $10^{-12}$. Therefore $r$ is virtually identical to
$R$ and Mr. Einstein's approximation is adequate to the
strongest requirements of the practice.\par
Finally, the exact form of the third Kepler's law for circular
orbits will be derived. Owing to (16) and (17), when one
sets $x=1/R$, for the angular velocity $n={d\phi/dt}$ it holds

$$n=cx^2(1-\alpha x).$$
For circular orbits both ${dx/d\phi}$ and ${d^2x/d\phi^2}$ must
vanish. Due to (18) this gives:

$$0={{1-h}\over c^2}+{{h\alpha}\over c^2}x-x^2+\alpha x^3,~~
0={{h\alpha}\over c^2}-2x+3\alpha x^2.$$
The elimination of $h$ from these two equations yields

$$\alpha=2c^2x(1-\alpha x)^2.$$
Hence it follows

$$n^2={\alpha\over 2}x^3={\alpha\over{2R^3}}={\alpha\over
{2(r^3+\alpha^3)}}.$$
The deviation of this formula from the third Kepler's law is
totally negligible down to the surface of the Sun. For an
ideal mass point, however, it follows that the angular velocity does
not, as with Newton's law, grow without limit when the radius of
the orbit gets smaller and smaller, but it approaches a determined
limit

$$n_0={1\over{\alpha\sqrt{2}}}.$$
(For a point with the solar mass the limit frequency will be
around $10^4$ per second). This circumstance could be of interest,
if analogous laws would rule the molecular forces.
\end